\documentclass[twocolumn,final]{elsarticle}
\usepackage{amssymb}
\usepackage{graphicx}
\journal{}

\def\nuc#1#2{\relax\ifmmode{}^{#1}{\protect\text{#2}}\else${}^{#1}$#2\fi}

\begin{document}
\begin{frontmatter}
\title{Discovery and Cross-Section Measurement of Neutron-Rich Isotopes in the Element Range from Neodymium to Platinum at the FRS}

\author[GSI]{J.~Kurcewicz \corref{cor1}}
\author[GSI,JLU]{F.~Farinon \corref{cor2}}
\author[GSI,JLU]{H.~Geissel}
\author[GSI]{S.~Pietri}
\author[GSI]{C.~Nociforo}
\author[GSI,JLU]{A.~Prochazka}
\author[GSI]{H.~Weick}
\author[GSI]{J.S.~Winfield}
\author[GSI,SMU]{A.~Estrad{\'e}}
\author[USP]{P.R.P.~Allegro}
\author[CEA]{A.~Bail}
\author[CEA]{G.~B{\'e}lier}
\author[USC]{J.~Benlliure}
\author[INFN]{G.~Benzoni}
\author[SUR]{M.~Bunce}
\author[SUR]{M.~Bowry}
\author[UPC]{R.~Caballero-Folch}
\author[GSI,JLU]{I.Dillmann}
\author[GSI,JLU,TOMSK]{A.~Evdokimov}
\author[GSI]{J.Gerl}
\author[LNL]{A.~Gottardo}
\author[GSI]{E.~Gregor}
\author[COM]{R.~Janik}
\author[GSI]{A.~Keli{\'c}-Heil}
\author[GSI]{R.~Kn\"obel}
\author[RIKEN]{T.~Kubo}
\author[GSI,MPI]{Yu.~A.~Litvinov}
\author[GSI,TUD]{E.~Merchan}
\author[GSI]{I.~Mukha}
\author[GSI,IKP]{F.~Naqvi}
\author[WAW]{M.~Pf\"utzner}
\author[WAW]{M.~Pomorski}
\author[SUR]{Zs.~Podoly{\'a}k}
\author[SUR]{P.H.~Regan}
\author[GSI,JLU]{B.~Riese}
\author[GSI]{M.V.~Ricciardi}
\author[GSI,JLU]{C.~Scheidenberger}
\author[COM]{B.~Sitar}
\author[GSI]{P.~Spiller}
\author[GSI]{J.~Stadlmann}
\author[COM]{P.~Strmen}
\author[JLU,PEK]{B.~Sun}
\author[COM]{I.~Szarka}
\author[CEA]{J.~Ta\"ieb}
\author[GSI,RIKEN]{S.~Terashima}
\author[LNL]{J.J.~Valiente-Dob\'on}
\author[GSI]{M.~Winkler}
\author[EDI]{Ph.~Woods}

\address[GSI]{GSI Helmholtzzentrum f\"{u}r Schwerionenforschung, 64291 Darmstadt, Germany}
\address[JLU]{Justus-Liebig-Universit\"{a}t Gie{\ss}en, 35392 Gie{\ss}en, Germany}
\address[SMU]{Astronomy and Physics Department, Saint Mary's University, Halifax, Nova Scotia B3H 3C3, Canada}
\address[USP]{Institute of Physics, Universidade de S\~{a}o Paulo, CEP 05508-090 Cidade Universit{\'a}ria, S\~{a}o Paulo, Brazil}
\address[CEA]{CEA DAM DiF, 91290 Arpajon Cedex, France}
\address[USC]{Universidad de Santiago de Compostela, E-15706 Santiago de Compostella, Spain}
\address[INFN]{INFN sezione di Milano, I-20133 Mialno, Italy}
\address[SUR]{Department of Physics, University of Surrey, Guildford, Surrey, UK GU2 7XH}
\address[UPC]{Universitat Polit{\`e}cnica de Catalunya, 08034 Barcelona, Spain}
\address[TOMSK]{National Research Tomsk Polytechnic University, Tomsk 634050, Russia}
\address[LNL]{INFN - Laboratori Nazionali di Legnaro, 35020 Legnaro, Italy}
\address[COM]{Faculty of Mathematics and Physics, Comenius University, 842 48 Bratislava, Slovakia}
\address[RIKEN]{RIKEN Nishina Center, RIKEN, Wako, Saitama 351-0198, Japan}
\address[MPI]{Max-Planck-Institut f\"ur Kernphysik, 69117 Heidelberg, Germany}
\address[TUD]{Institut f\"ur Kernphysik, Technische Universit\"at Darmstadt, 62289 Darmstadt, Germany}
\address[IKP]{Institut f{\"u}r Kernphysik,Universit\"{a}t zu K\"oln, 50937 K\"oln, Germany}
\address[IEP]{Faculty of Physics, University of Warsaw, 00-681 Warsaw, Poland}
\address[PEK]{School of Physics and Nuclear Energy Engineering, Beihang University, Beijing 100191, China}
\address[EDI]{School of Physics, University of Edinburgh, Edinburgh EH9 3JZ, UK}
\cortext[cor1]{Corresponding author; electronic address: j.kurcewicz@gsi.de}%
\cortext[cor2]{Part of PhD work, Justus-Liebig University, Gie{\ss}en, 2011}%

\begin{abstract}
With a new detector setup and the high-resolution performance of the
fragment separator FRS at GSI we discovered 57 new isotopes in the
atomic number range of 60$\leq Z \leq 78$: \nuc{159-161}{Nb},
\nuc{160-163}{Pm}, \nuc{163-166}Sm, \nuc{167-168}{Eu},
\nuc{167-171}{Gd}, \nuc{169-171}{Tb}, \nuc{171-174}{Dy},
\nuc{173-176}{Ho}, \nuc{176-178}{Er}, \nuc{178-181}{Tm},
\nuc{183-185}{Yb}, \nuc{187-188}{Lu}, \nuc{191}{Hf},
\nuc{193-194}{Ta}, \nuc{196-197}{W}, \nuc{199-200}{Re},
\nuc{201-203}{Os}, \nuc{204-205}{Ir} and \nuc{206-209}{Pt}. The new
isotopes have been unambiguously identified in reactions with a
$^{238}$U beam impinging on a Be target at  1 GeV/u. The isotopic
production cross-section for the new isotopes have been measured and
compared with predictions of different model calculations. In
general, the ABRABLA and COFRA models agree better than a factor of
two with the new data, whereas the semiempirical EPAX model deviates
much more. Projectile fragmentation is the dominant reaction
creating the new isotopes, whereas fission contributes significantly
only up to about the element holmium.

\end{abstract}
\begin{keyword}
NUCLEAR REACTIONS Be(\nuc{238}{U},$X$), $E = 1$ GeV/u; measured Nb-Pt fragments isotopic
production $\sigma$; Comparison with previous results and model predictions.
%
\end{keyword}
%

\end{frontmatter}

\section{Introduction}

Heavy neutron-rich nuclides are of great interest for nuclear astrophysics and
basic nuclear spectroscopy. This
becomes immediately obvious when one looks at the predicted path for
r-process nuclei and their decay. The study of shell evolution far
off stability and towards the expected magic numbers $N$=82, 126 and
thus the waiting points of the r-process nuclides are of interest
for both fields.
The accurate knowledge of the atomic masses and  lifetimes are
essential for the understanding of the nucleosynthesis
\cite{KTW98,FGM08,Roe09}. Presently,  the corresponding theories
when applied to newly opened experimental territories still deviate
significantly from the results of measurements.

Experimentally the area of heavy neutron-rich nuclides is difficult
to reach because of the low production cross-sections and the great
challenge of separation and isotopic identification. Relativistic
energies of the reaction products and the high ion-optical
resolution of the in-flight separator FRS \cite{frs} are the keys to
the frontiers in this domain of nuclides. High velocities are
required to reduce the number of populated ionic charge-states for
each element, mainly to bare fragments with a low contamination of
H- and He-like ions.

Several milestones in nuclear physics have been achieved with the
FRS, like the discovery and spectroscopy of the double magic
nuclei \nuc{100}{Sn} \cite{schneider,hin10} or the discovery of 2p
radioactivity in \nuc{45}{Fe} \cite{pfutzner}.  In the pioneering
experiments with uranium projectile fission with  the FRS at 750
MeV/u more than 120 new isotopes have been discovered, among them
\nuc{78}{Ni} \cite{eng95}. These achievements have launched a new
research activity for fission studies \cite{SSB00} and an area of
applied physics towards accelerator-driven reactors and
nuclear-waste transmutation \cite{Ben99}. Owing to the success of
in-flight fission at high energies, all next-generation in-flight
exotic nuclear beam facilities include the production via
projectile fission. For example, very recently the new powerful
Radioactive Ion Beam Factory (RIBF) in RIKEN has successfully
started the experimental program with the discovery of new
isotopes in the atomic number range of $26\leq Z \leq 56$ produced
via in-flight fission of an intense \nuc{238}{U} beam
\cite{ohn10}. The search for the neutron dripline at low $Z$ has
been a major research activity at GANIL (France)\cite{Luk02}, NSCL
(USA)\cite{Tar09}  and RIKEN (Japan)\cite{Sak99,Luk02} for many
years. At the NSCL facility at MSU (USA) new neutron-rich isotopes
have been observed via reactions with a \nuc{48}{Ca} and
\nuc{76}{Ge}  beam at about 140 MeV/u \cite{Bau07,Tar09}.

In the recent years, the intensity for \nuc{238}{U} beams provided
by the GSI accelerators have increased almost by a factor 10, which
has opened new perspectives for the production and study of the
heaviest projectile fragments \cite{alv10, che10}. Even along with
mass measurements at the FRS-ESR facility new isotopes have been
observed \cite{RNB6-HG,che10}. The fragmentation reaction of
\nuc{208}{Pb} seems also to be very promising for production of
neutron-rich isotopes as it has been proven in the recent FRS
experiments \cite{kur06,kur10}.

In this letter we report on the discovery of 57 new neutron-rich
isotopes in the element range of Nd to Pt at the FRS by the use of
new particle identification methods \cite{Far11}.

\section{Experimental technique}

\begin{figure}[h!]
\begin{center}
\includegraphics[width=0.9\columnwidth]{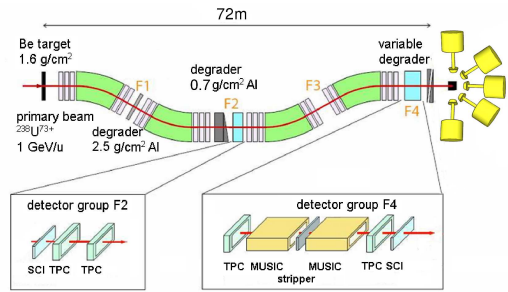}
\caption{(color online) Schematic view of the FRS and the
experimental setup. The magnetic dipole and quadrupole magnets of
the FRS and the target and different focal-plane areas are depicted.
The different energy degraders at the focal planes F1, F2, and F4
are schematically shown. The detector setups placed in the central
and final focal planes are shown as zoomed inserts in the picture.
After identification in flight the ions are stopped in an
implantation detector viewed by a gamma detector array. }
\label{fig:frs}
\end{center}
\end{figure}

The experiment was performed with the SIS-18 synchrotron of GSI Darmstadt, which
delivered a 1 GeV/u \nuc{238}{U} beam in spills lasting 0.5-2~s with
a repetition period of 2-4~s. The beam impinged
on a 1.6 g/cm$^2$ thick beryllium target placed at the entrance of the
projectile FRragment Separator (FRS) \cite{frs}.
The primary beam intensity was of the order of  $2\times 10^{9}$ ions/spill.
The \nuc{238}{U} intensity was recorded
by a calibrated secondary-electron transmission monitor~\cite{seetram}.
The reaction products were separated by the FRS operated in an
overall achromatic ion-optical mode. A schematic view of the FRS and
the experimental setup is shown in Fig.~\ref{fig:frs}. The spatial
separation in flight was achieved by twofold application of the $B
\rho$-$\Delta E$-$B \rho$  method, i.e., the atomic energy losses in
two degraders, located at the first (F1) and second (F2) focal
planes, were measured via magnetic rigidity analysis. In this way,
the reaction products are spatially separated and by the use of
various detectors their nuclear charge $Z$ and mass number $A$ could
be determined. After the first magnetic selection and the 2.5
g/cm$^2$ thick aluminum degrader at F1, the reaction products were
slowed down in an aluminium disk degrader located at the
intermediate focal plane F2. With the disk angle the degrader shape
was tuned to preserve the achromatism. Even at these relativistic
velocities the atomic charge states of the heavy fragments represent
a separation problem. Therefore, medium $Z$-material niobium foils
were placed both behind the target and the F2-degrader to enhance
the yield of bare fragments. The thicknesses of these electron
strippers were 223~mg/cm$^2$ for the first and 106 mg/cm$^2$ for the
latter. The total thickness of the F2-materials including detectors
was 1.43~g/cm$^2$ aluminium equivalent.

The complete particle identification in-flight was performed on an
event-by-event basis with time-of-flight (ToF), energy-deposition
($\Delta E'$) and magnetic rigidity measurements (ToF-$\Delta
E'$-$B\rho$ method). The ToF measurement was performed with two
plastic scintillator detectors, one located at F2 and the other one
at the final focal plane (F4). The flight path was about 37 m
between the two detectors. The ToF value for the selected isotopes
was of the order of 160 ns in the laboratory frame. At the exit of
the FRS, two ionization chambers (MUSIC) filled with P10 gas at 1
atm. pressure were mounted  with a 104 mg/cm$^2$ copper stripper
placed in between. The MUSIC detectors  \cite{music} delivered the
energy-deposition  signals of fragments, thus providing the
information of their atomic numbers. The velocity dependence of the
penetrating ions was taken into account before the energy-deposition
signal was applied for $Z$ identification. The magnetic rigidity
measurements were performed with four time-projection chambers
(TPC), two located at the dispersive focal plane (F2) and two others
mounted at the exit of the FRS. The TPC provided full tracking
information (angle and position) for the transmitted fragments. The
event-by-event identification, and thus also the in-flight
separation, were verified by the isomer tagging technique
\cite{Far11}. In the range of the particle identification spectrum,
known $\mu$s isomers were selected (\nuc{171m}{Tm}, \nuc{172m}{Yb}
and \nuc{175m}{Lu}) whose gamma rays were recorded in coincidence
with the incoming ions. With these multiple redundant measurements
and the two-stage in-flight separation criteria, we achieved an
unambiguous isotope identification. Finally, a range selection was
applied  in addition because the selected fragments were stopped in
a layer of matter viewed either  by the RISING germanium detector
setup \cite{pietri}, composed of fifteen Euroball cluster of seven
crystals used in 4$\pi$ configuration, or the simpler isomer tagging
device \cite{Far11}, which consisted of the two electro-mechanically
cooled  Ge detectors, a stopper foil,
and a veto scintillation counter. 

\section{Data analysis}

In the experiment four different $B\rho$ settings of the FRS were
applied, which were chosen to yield optimum  intensities for bare
\nuc{172}{Dy}, \nuc{194}{Os}, \nuc{198}{Os} and \nuc{202}{Os} ions,
i.e., the magnetic fields were selected to center these isotopes
during their travel inside the FRS system. The data collected in
each setting were processed by the following procedure based on a
combination of ToF, position and energy-deposition.

In the first step of analysis a two-dimensional plot of the energy
deposition in the first MUSIC detector as a function of the energy
deposition signal from the second MUSIC was created. Only bare ions
in both detectors were chosen for further analysis, thus rejecting
ions with different charge states in both MUSICs or secondary
reaction products created in the detector material.

In the second step, a distribution of the reconstructed nuclear
charge as a function of the energy loss of the ions in the
degrader located at
F2 was created.  This allowed a clear identification of the group of ions which did not change their charge state while penetrating
the matter placed at F2.

\begin{figure}[t!]
\begin{center}
\includegraphics[width=1.0\columnwidth]{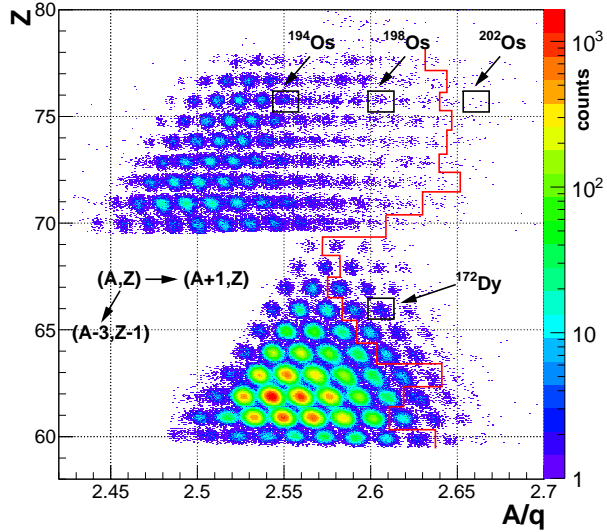}
\caption{(color online) Identified atomic number $Z$ of the incoming
ions as a function of their $A/q$ ratio, at the final focal plane
(F4). In the $A/q$ ratio,  the time-of-flight and magnetic rigidity
analysis information is included. The plot shows the superimposed
data acquired in all $B\rho$ settings.  The solid line shows the
border of hitherto unobserved isotopes, i.e., the discovered new
isotopes are on the right hand side of the border line. The black
rectangles show the group of ions corresponding to the $B\rho$
setting of the spectrometer. } \label{fig:id}
\end{center}
\end{figure}

\begin{figure*}
\begin{center}
\includegraphics[width=1.0\textwidth]{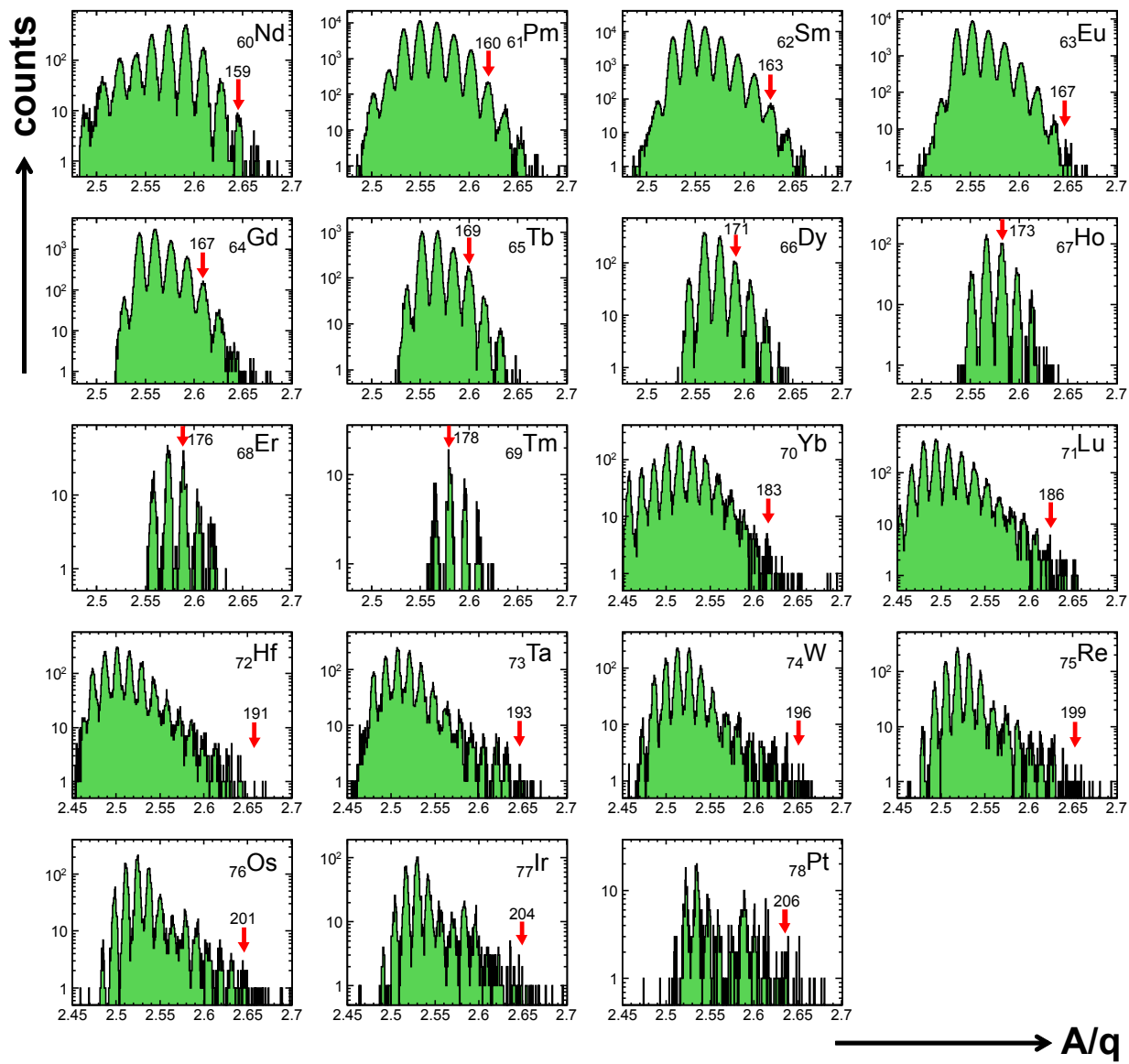}
\caption{(color online) Projection of the identification plot
shown on Fig.~\ref{fig:id} constructed for all elements covered by
the different  $B\rho$ settings of the FRS. For each element the
arrow indicates the lightest of the isotopes (marked by the mass
numbers) observed for the first time. }\label{fig:proj}
\end{center}
\end{figure*}

Unambiguous isotope identification  requires an additional selection
based on the ions' position at the final focus \textbf{($B \rho$)},
energy deposition signals in both scintillators and a correlation
between the measured angles at the intermediate and final focal
planes. The latter condition reflects the ion-optical image
conditions which are needed also for the correct $B \rho$
determination.
An example of the final isotope identification plot is shown in
Fig.~\ref{fig:id}. The projection of this plot on the $A/q$ axis
selecting every element covered in the $B\rho$ settings is given in
Fig.~\ref{fig:proj}
\section{Production cross-sections}

The data recorded for each $B \rho$ setting were analyzed by the
described procedure to achieve an unambiguous isotope
identification. In the next step the production cross-sections of
individual  isotopes  were determined according to:
\begin{equation}
 \sigma_f = \frac{N_f}{T_{opt} T_{sec} P_0 Y N_p f_{dt}},
\label{eq:sigma}
\end{equation}
where $N_f$ is the number of registered ions of a certain isotope,
$T_{opt}$ the ion-optical transmission, $T_{sec}$ a correction
for secondary reactions in the matter placed after the target (e.g., degraders and detectors), $P_{0}$ the probability that an ion remains fully
stripped in both stages of the separator,
$Y$ the correction for losses of primary beam and fragments due to nuclear
reactions in the target material, $N_p$ the total number of \nuc{238}{U}
ions and  $f_{dt}$ the correction for the dead-time losses of the data acquisition
system.
All secondary reactions in the target and the  matter in the focal planes are taken into account
by applying the Benesh-Cook-Vary formula \cite{benesh}.

Calculations of charge-state distributions were performed for the
heavy fragments of interest  with the CHARGE code~\cite{charge}
yielding $P_0$ values in the range of 0.9 and 0.6 for Nd-Pt
isotopes, respectively. The ion-optical transmissions $T_{opt}$ have
been calculated by using the Monte-Carlo simulation program MOCADI
\cite{mocadi,mocadi_code} assuming the kinematics of projectile
fragmentation \cite{Morrissey}.  The transmission values were
obtained separately for each $B\rho$ setting. The typical $T_{opt}$
values  were of the order of 0.4-0.6 for isotopes with $A/q$  close
to the reference setting of each FRS setting. This relative small
value is due to the tight slits settings of the FRS which were
applied in order to decrease the number of contaminants, mainly
light fission fragments (Z$<$60), reaching the F4 area.

\section{Results}

In this experiment  we discovered 57 new isotopes with atomic
numbers in the range 60$\leq Z \leq 78$: \nuc{159-161}{Nb},
\nuc{160-163}{Pm}, \nuc{163-166}Sm, \nuc{167-168}{Eu},
\nuc{167-171}{Gd}, \nuc{169-171}{Tb}, \nuc{171-174}{Dy},
\nuc{173-176}{Ho}, \nuc{176-178}{Er}, \nuc{178-181}{Tm},
\nuc{183-185}{Yb}, \nuc{187-188}{Lu}, \nuc{191}{Hf},
\nuc{193-194}{Ta}, \nuc{196-197}{W}, \nuc{199-200}{Re},
\nuc{201-203}{Os}, \nuc{204-205}{Ir} and \nuc{206-209}{Pt} and
measured their production cross section.  The new isotopes have been
unambiguously identified in flight by applying a two-fold $B
\rho$-$\Delta E$-$B\rho$ separation scenario and redundant
ToF-$\Delta E'$-$B \rho$ analysis using new detector systems
\cite{Far11}. The observed new isotopes are presented for the
different elements in Fig.~\ref{fig:proj}. For each element the
arrow indicates the lightest of the isotopes observed for the first
time. For all isotopes discovered in this work we can set a lower
limit of half-life to 300 ns which corresponds to the time of flight
between the production target and F4.

\begin{figure*}
\begin{center}
\includegraphics[width=1.0\textwidth]{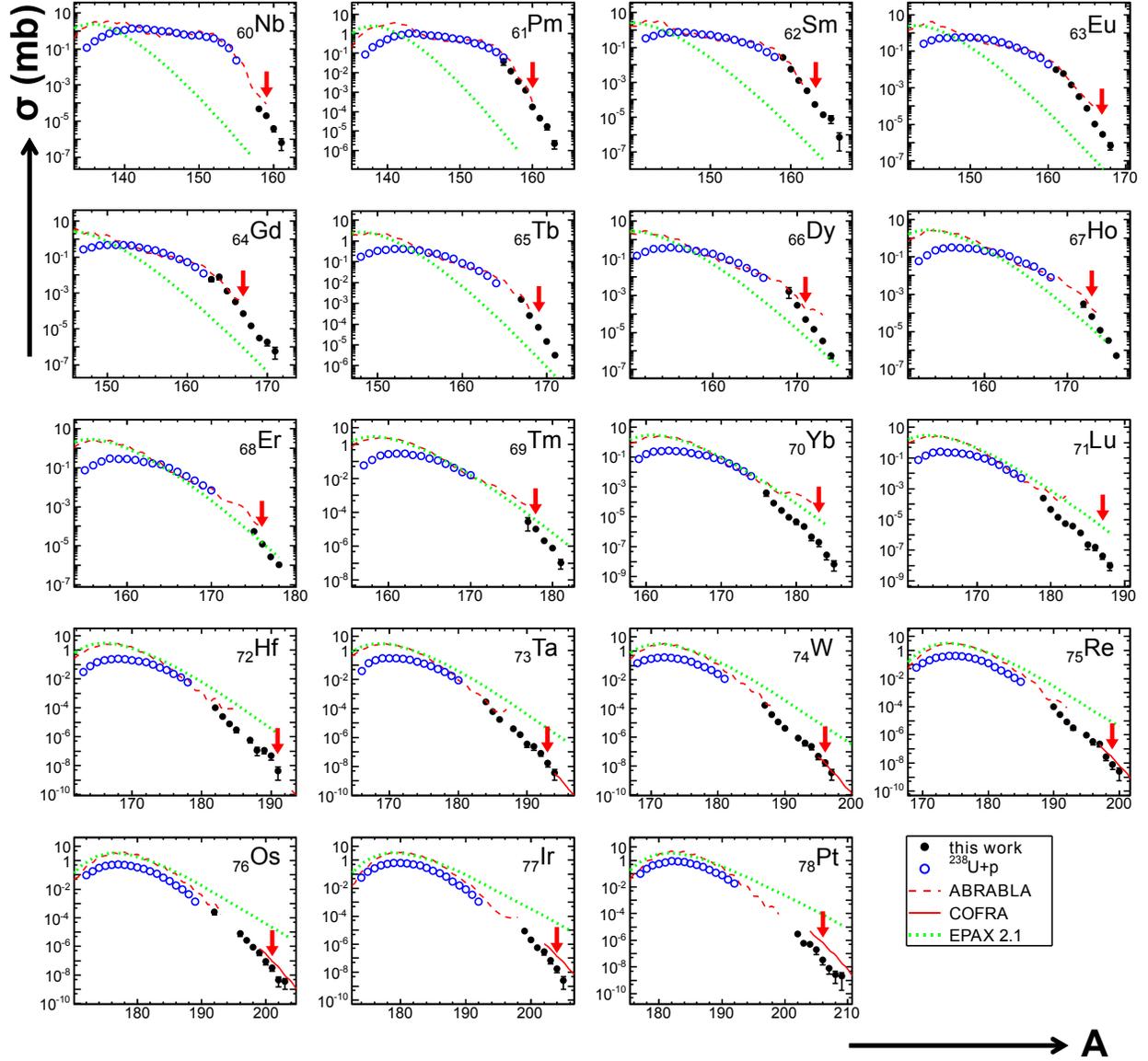}
\caption{(color online) Measured production cross-sections of
fragments produced in the reaction \nuc{238}{U} (1 GeV/u) + Be
(black circles), shown together with the experimental results
obtained by Bernas et al. \cite{bernasfi} ($Z=60-64$), \cite{bernas}
($Z=65-73$) and Ta{\"i}eb et al. \cite{taieb} ($Z=74-78$)
 in the reaction
\nuc{238}{U} (1 GeV/u) + p (blue open circles). The red dashed line
represents the predictions of the ABRABLA model \cite{gaimard} and
the continuous red line shows the results of COFRA \cite{cofra}
($Z=73-78$). The green dotted line shows the prediction of EPAX
model \cite{epax}. For each element the arrow indicates the lightest
of the isotopes observed for the first time.
}\label{fig:xsec}
\end{center}
\end{figure*}

The isotopic production cross-section have been determined in this
experiment down to the pico-barn level. Particularly, the steep
descent of the yields of the neutron-rich isotopes has been mapped.
The cross-section obtained for the production of Nd-Pt isotopes are
shown in Fig.~\ref{fig:xsec} and listed in Table~\ref{tab:xsec}. The
error bars are mainly determined by the statistics due to tiny
production cross-sections, partially due to the low transmission of
isotopes characterized with a B$\rho$ value far from the central
field setting with respect to the optical axis. The latter aspect
has been partially taken care of by the different field settings
applied in these measurements. The measured cross-sections in this
experiment represent the outskirts of the chart of nuclides.
Therefore, it is obvious that the present results cannot be compared
with previous experimental data. However, a good orientation for the
consistencies of the cross-section evolution can be obtained with
the
 comparison to results from previous experimental studies
 \cite{bernas, taieb}. Although the latter data from the literature are
obtained from reactions of 1 GeV/u \nuc{238}{U} beam with a liquid
hydrogen target the continuous transition of the two experimental
data sets is remarkable. The excitation in the reaction with
hydrogen target nuclei should be lower than in our case with a
beryllium target but this plays  obviously a minor role for the
compared fragment distribution in the overlap region. Our
cross-section are compared with calculations based on the ABRABLA
\cite{gaimard} and COFRA \cite{cofra} models and shown in
Fig.~\ref{fig:xsec}. The ABRABLA model is an improved
abrasion-ablation model which takes also into account microscopic
structural effects and the contribution of fission.  COFRA and EPAX
have both the fission process not included. The predictions of the
semiempirical EPAX parametrization \cite{epax} are given as well.

Since ABRABLA is a Monte-Carlo code,  a long computation time is
required to  reach very-low production cross-section. Thus COFRA,
the analytical version of the abrasion-ablation model, was used for
the most neutron-rich nuclei. The predictions of ABRABLA show an
overall good agreement with the experimental data. The used version
of the code involves fission as a possible deexcitation step of the
created prefragment. The lighter isotopes in the element range Nb-Dy
show a significant contribution from the fission process which
starts to play a minor role for heavier nuclei investigated in this
work. The latter statement was deduced from the ABRABLA calculation
and is also visible by the underestimation of EPAX, which takes into
account only projectile fragmentation.

The COFRA model agrees well with the experimental data, however, a
small gradual deviation between results of the measurements and the
model predictions can be observed for the heaviest (Os-Pt) nuclei.
For most of the cases the overestimation is less than a factor of
two compared with the measured values which means it is well suited
to make valid predictions for new studies in this field of
neutron-rich isotopes. The EPAX predictions largely overestimates
the production cross-sections in the region of neutron-rich nuclei
in the element range of Yb to Pt. This reflects the present
limitation of that semiempirical model with its parametrization
based on previous data, hence our results could be a base for a new
parametrization of EPAX.

\begin{table*}
\begin{center}
\caption{Production cross-section of new isotopes measured in this work in reaction 1 GeV/u \nuc{238}{U} on Be.}
\label{tab:xsec}
\begin{tabular}{cccccccc}
\hline
Isotope & $\sigma$(nb) & Isotope & $\sigma$(nb) & Isotope & $\sigma$(nb) & Isotope & $\sigma$(nb) \\
\hline
 \nuc{159}{Nb} & 20(6) & \nuc{169}{Gd} & 3.2(5) & \nuc{177}{Er} & 2.7(0.3) & \nuc{197}{W} & 0.0034(17) \\
 \nuc{160}{Nb} & 3.9(1.4) & \nuc{170}{Gd} & 1.9(8) & \nuc{178}{Er} & 1.1(2) & \nuc{199}{Re} & 0.0076(27) \\
 \nuc{161}{Nb} & 0.7(4) & \nuc{171}{Gd} & 0.6(4) & \nuc{178}{Tm} & 10.6(1.8) & \nuc{200}{Re} & 0.0025(14) \\
 \nuc{160}{Pm} & 180(15) & \nuc{169}{Tb} & 71(4) & \nuc{179}{Tm} & 2.1(3) & \nuc{201}{Os} & 0.033(7) \\
 \nuc{161}{Pm} & 45(6) & \nuc{170}{Tb} & 14.6(1.1) & \nuc{180}{Tm} & 0.79(17) & \nuc{202}{Os} & 0.0044(2) \\
 \nuc{162}{Pm} & 17(6) & \nuc{171}{Tb} & 3.3(0.4) & \nuc{181}{Tm} & 0.11(6) & \nuc{203}{Os} & 0.0035(17) \\
 \nuc{163}{Pm} & 2.2(1.0) & \nuc{171}{Dy} & 49(3) & \nuc{183}{Yb} & 0.21(6) & \nuc{204}{Ir} & 0.017(5) \\
 \nuc{163}{Sm} & 53(4) & \nuc{172}{Dy} & 15.0(1.1) & \nuc{184}{Yb} & 0.028(9) & \nuc{205}{Ir} & 0.0026(15) \\
 \nuc{164}{Sm} & 14(2) & \nuc{173}{Dy} & 3.5(4) & \nuc{185}{Yb} & 0.007(3) & \nuc{206}{Pt} & 0.033(11) \\
 \nuc{165}{Sm} & 8.3(4) & \nuc{174}{Dy} & 0.57(18) & \nuc{187}{Lu} & 0.043(9) & \nuc{207}{Pt} & 0.008(3) \\
 \nuc{166}{Sm} & 0.7(5) & \nuc{173}{Ho} & 65(6) & \nuc{188}{Lu} & 0.010(3) & \nuc{208}{Pt} & 0.0027(15) \\
 \nuc{167}{Eu} & 2.9(7) & \nuc{174}{Ho} & 12(1) & \nuc{191}{Hf} & 0.0043(25) & \nuc{209}{Pt} & 0.0020(14) \\
 \nuc{168}{Eu} & 0.7(3) & \nuc{175}{Ho} & 3.4(4) & \nuc{193}{Ta} & 0.017(5) &  &  \\
 \nuc{167}{Gd} & 71(4) & \nuc{176}{Ho} & 0.51(14) & \nuc{194}{Ta} & 0.0037(19) &  & \\
 \nuc{168}{Gd} & 14.9(1.2) & \nuc{176}{Er} & 1.19(12) & \nuc{196}{W} & 0.018(4) &  &  \\
\hline
\end{tabular}
\end{center}
\end{table*}

\section{Summary}
The results of the present experiment opens a new field for nuclear
spectroscopy and also for nuclear astrophysics in the heavy nuclei
range. With a new detector setup and the high-resolution performance
of the fragment separator FRS  we discovered 57 new isotopes in the
atomic number range of $60 \leq Z \leq 79$. The new isotopes have
been unambiguously identified in projectile fragmentation reactions
with a $^{238}$U beam at  1 GeV/u. The isotopic production
cross-section for the new isotopes have been measured and compared
with the predictions of the ABRABLA, COFRA, and EPAX models. In
general, with COFRA a good overall agreement has been achieved to be
confident for reliable predictions in unknown territory. The next
steps in this experimental campaign will be half-life and mass
measurements, as well as decay spectroscopy after implantation in
silicon detectors.

\section{Acknowledgement}

It is a pleasure to thank the technical staffs of the
accelerators, the FRS, and the target laboratory for their
valuable contribution to the beam quality and experimental setups.
The authors gratefully acknowledge fruitful discussions with
G. Martinez-Pinedo, K. Otsuki and B. Pfeiffer.
This work was supported by STFC (UK). B.Sun is partially supported by NECT and NSFC 11105010.

\end{document}